\documentclass[adp,fleqn]{w-art}
\usepackage{times}
\usepackage[]{graphicx}

\usepackage{amssymb}
\usepackage{amsmath,bm}

\def\a{\alpha}
\def\b{\beta}

\begin{document}

\Dateposted{{\it
    file ChargeNoScrew18x.tex}}
\keywords{Maxwell's equations, polar and axial vectors, differential
  forms with twist, premetric electrodynamics}
\subjclass[pacs]{03.50.De, 06.30.Ka, 01.65.+g}
\title[An electric charge has no screw sense]{An electric charge has
  no screw sense---a comment on the twistfree formulation of
  electrodynamics by da Rocha \& Rodrigues}
\author[Y.~Itin]{Yakov Itin\inst{1,}%
  \footnote{Corresponding author $\,$ E-mail:~\textsf{itin@math.huji.ac.il}} }
\address[\inst{1}]{Institute of Mathematics, The Hebrew University of
  Jerusalem\\{\it and}\\ Jerusalem College of Technology, Jerusalem,
  Israel}
\author[Yu.N.~Obukhov]{Yuri
  N.~Obukhov\inst{2,}\footnote{E-mail:~\textsf{yo@thp.uni-koeln.de}}}
\address[\inst{2}]{Department of Mathematics, University College
  London, Gower Street, London WC1E 6BT, UK}
\author[F.W.~Hehl]{Friedrich
  W.~Hehl\inst{3,}\footnote{E-mail:~\textsf{hehl@thp.uni-koeln.de}} }
\address[\inst{3}]{Institute for Theoretical Physics, University of
  Cologne, 50923 K\"oln, Germany   \\{\it and} \\
  Department of Physics and Astronomy, University of
  Missouri, Columbia, MO 65211, USA}

\begin{abstract}
Da Rocha and Rodigues (RR) claim (i) that in classical electrodynamics
in {\it vector calculus} the distinction between polar and axial
vectors and in {\it exterior calculus} between twisted and untwisted
forms is inappropriate and superfluous, and (ii) that they can derive
the Lorentz force equation from Maxwell's equations. As to (i), we
point out that the distinction of polar/axial and twisted/untwisted
derives from the property of the electric charge of being a pure
scalar, that is, not carrying any screw sense. Therefore, the
mentioned distinctions are necessary ingredients in any fundamental
theory of electrodynamics. If one restricted the allowed coordinate
transformations to those with positive Jacobian determinants (or
prescribed an equivalent constraint), then the RR scheme could be
accommodated; however, such a restriction is illegal since
electrodynamics is, in fact, also covariant under transformations with
negative Jacobians. As to (ii), the ``derivation'' of the Lorentz force
{}from Maxwell's equations, we point out that RR forgot to give the
symbol $F$ (the field strength) in Maxwell's equations an operational
meaning in the first place. Thus, their proof is empty. Summing up:
the approach of RR does not bring in any new insight into the
structure of electrodynamics.

\end{abstract} 
\maketitle 


This paper is a reaction to some claims of da Rocha \& Rodrigues
\cite{dR&R} related to classical electrodynamics. For this purpose we
begin with a brief and rough sketch of how the modern premetric form
of Maxwell's equations came about and how the premetric framework is
based on an appropriate operational interpretation.

\section{Maxwell's equations in space and time}

\subsection{In components}

Maxwell \cite{Maxwell} formulated his equations in terms of Cartesian
components. If we use Cartesian coordinates $x^a$, with
$a,b,..=1,2,3$, and the time $t=x^0$, then we have the inhomogeneous
and the homogeneous Maxwell equations as, respectively,
\begin{eqnarray}\label{Inhom}
\partial_1 D_1+\partial_2 D_2+\partial_3 D_3=\rho\,,&&\qquad
\partial_2 H_3-\partial_3 H_2-\partial_0 D_1=j_1 \quad\text{and cyclic}\,,\\
\partial_1 B_1+\partial_2 B_2+\partial_3 B_3=0\,,&&\qquad\hspace{3pt}
\partial_2 E_3-\partial_3 E_2-\partial_0 B_1=0\hspace{4pt} \quad
\text{and cyclic}\,.\label{Hom}
\end{eqnarray}
Clearly, in this framework with its so-called Cartesian vectors, we
don't need to distinguish between upper and lower indices, nor talk
about densities. A screw sense is naturally defined by the sequence
$x^1,x^2,x^3$ of the Cartesian axes. The quantities $D,H;B,E;j$ are
all 3-dimensional (3d) vectors, as mathematical objects they are all
alike. Accordingly, the concept of a polar and an axial vector does
not exist in electrodynamics as long as we restrict ourself to proper
rotations $SO(3)$, that is, as long as the 3d Jacobian
$J_3:=\text{det}\,||\partial x^a/\partial x^{a'}\!||$ is positive:
$J_3=+1$. However, this restriction is unphysical since we know that
Mawell's equations are, in fact, covariant under improper
transformations, too.

With hindsight we know that the general 3-dimensional
orthogonal group $O(3)$ is a symmetry group of the field equations
(\ref{Inhom}) and (\ref{Hom}), and $D,H;B,E;j$ are vector
representations of $SO(3)$ and $\rho$ a scalar representation
therefrom.

\subsection{In vector calculus}

Curious as mankind is, one didn't want to restrict oneself to proper
transformations, that is, space reflections should be included, and at
the same time a transition to curvilinear coordinates was
desirable. As a nice formulation, the vector form of the Maxwell
equations came up around the turn of the 19th to the 20th century, see
Abraham \& F\"oppl \cite{Abraham}. In the form given by Jackson
\cite{Jackson}, they read
\begin{eqnarray}  \label{Inhom'} 
\text{div}\,{\mathbf D}=\rho\,,\qquad&& \text{curl}\,{\mathbf H} 
  - \frac{\partial{\mathbf D}}{\partial t}={\mathbf j}\,,\\
\label{Hom'}
\text{div}\,{\mathbf B}=0\,,\qquad&& \hspace{3pt}\text{curl}\,{\mathbf E} 
+ \frac{\partial{\mathbf B}}{\partial t}=0\,.
\end{eqnarray}
An integral part of electrodynamics is the equation that defines
$\mathbf{E}$ and $\mathbf{B}$ in the first place, namely the
expression of the Lorentz force 
\begin{equation}\label{Lor} {\mathbf F}=q\left({\mathbf E}+{\mathbf
      v}\times {\mathbf B}\right)\,.
\end{equation}

If we make a space reflection $x^a\longrightarrow -x^a$, we want that
(\ref{Lor}) stays invariant. If we write (\ref{Lor}) in components,
\begin{equation}\label{Lor'}
  F^1=q(E^1+v^2 B^3-v^3 B^2)\quad\text{etc.},
\end{equation}
we immediately recognize that in the product $v^2 B^3$ only one vector
component can turn around its sign upon reflection. Since the velocity
$\mathbf{v}$ is the prototype of a (contravariant) vector, it must be
$\mathbf{B}$ that is promoted to an axial vector that remains
invariant under reflections. 

This knowledge applied to (\ref{Hom'})$_2$, uncovers the curl operator
as axial vector. In (\ref{Inhom'})$_2$, since $\mathbf{j}$ is polar
because of $\mathbf{j}=\rho\mathbf{v}$, with a scalar $\rho$, the
magnetic excitation $\mathbf{H}$ is recognized as axial and the
electric excitation $\mathbf{D}$ as polar. Accordingly, in this
context, $\mathbf{E},\mathbf{D}$ are polar and $\mathbf{H},\mathbf{B}$
axial vectors.

RR claim, see last phrase of their abstract, that ``We recall also a
formulation of the engineering version of Maxwell equations using
electric and magnetic fields as objects of the same nature, i.e.,
without using polar and axial vectors.'' We do, too, see equations
(\ref{Inhom}) and (\ref{Hom}). However, then, in a Cartesian calculus,
they have to require $J_3>0$. The more the symmetry group of a
physical system is widened, the more refined the description becomes
of the quantities involved. In contrast to RR, we believe that the
property of a vector being axial or polar is observable, one just has
to apply a space reflection in an electrostatic or in a induction
experiment, respectively.  Accordingly, RR want to widen the
transformation group---after all they work in arbitrary
coordinates---but they do not want to use the more refined description
of the field quantities involved.

In order to understand the procedure of RR better, we made another
attempt, compare also Gelman \cite{Gelman} and Brevik
\cite{Brevik}. We introduced a constant pseudoscalar field
$\beta$. Then, the polar fields $\widetilde{\mathbf{H}}:=\b\,
\mathbf{H}$ and $\widetilde{\mathbf{B}}:=\b\, \mathbf{B}$ can be
introduced and Maxwell's equations rewritten in terms of the polar
fields
$\mathbf{E},\mathbf{D},\widetilde{\mathbf{H}},\widetilde{\mathbf{B}}$. If
one wants to preserve the meaning of the differential operators div
and curl, we have to require $\beta^2=1$, that is, $\b=\pm 1$, with
the positive sign for a chosen orientation and the negative sign for
an opposite orientation. To achieve consistency, we have to redefine
the cross product and the curl operator by multiplying each with
$\b$. Formally, this can be done. However, one runs into all sorts of
strange behavior.  We arrive at a modified determinant with rather
curious properties: Its sign changes under a permutation of rows but
does not change under a permutation of columns (which is a
transformation of coordinates with a negative Jacobian). Moreover,
this modified determinant must remain invariant under multiplication
of its columns by $(-1)$. Accordingly, most properties of determinants
are lost. Even worse, the {\it mass} of a body will then necessarily
become negative in response to the change of the orientation, as it is
demonstrated in Sec. 6.1 of \cite{dR&R}. In a domino effect, the
elastic stress will become orientation-dependent, too \cite{Yavari}.
This also will apply to the classical action.  Why should we redefine
the vector product, handle strange ``determinants'' and negative
masses in order to be able to follow RR on their adventurous journey
to Absurdistan? We prefer to stick with polar and axial vectors and
follow the usual rationale of vector calculus.

Whatever the constructions of RR may mean, they certainly do not yield
a simpler representation of electrodynamics.

\subsection{In tensor calculus}

One could ask, why should we turn to tensor calculus, see Schouten
\cite{Schouten*}, if the vector calculus works so well. There are two
reasons: (i) the transition to arbitrary coordinates is more smooth,
(ii) the transition to spacetime is more smooth; hence we catch two
flies at once.

In vector calculus the operators div and curl take a fairly
complicated form in curvilinear coordinates. It is desirable to
circumvent this complication. The technique is well known: One
introduces $\rho$ as scalar {\it density}, we call it $\hat{\rho}$,
and $\mathbf{j}$ as contravariant vector {\it density} $\hat{j}^a$. As a
consequence also the electric excitation $\mathbf{D}$ becomes a
density $\mathfrak{D^a}$; densities will be printed in fracture style
or with a hat. The divergence operator then translates into
div$\,\mathbf{D}$ $\rightarrow\partial_a\mathfrak{D}^a$ and the curl
into curl$\,\mathbf{E}$ $\rightarrow\partial_a E_b-\partial_b E_a$;
the new operators are covariant under arbitrary coordinate
transformations.  Accordingly, the tensor version of (\ref{Inhom'})
and (\ref{Hom'}) reads
\begin{eqnarray}\label{Inhom''}
  \partial_1 \mathfrak{D}^1+\partial_2  \mathfrak{D}^2+\partial_3  
\mathfrak{D}^3= \hat{\rho}\,,&&\qquad
\partial_2 H_3- \partial_3 H_2  -\partial_0  \mathfrak{D}^1=\hat{j}^1 
\quad\text{and cyclic}\,,\\
\partial_1  \mathfrak{B}^1+\partial_2  \mathfrak{B}^2+\partial_3 
 \mathfrak{B}^3=0\,,&&\qquad
\partial_2 E_3-\partial_3 E_2  +\partial_0  \mathfrak{B}^1=0\hspace{4pt} \quad
\text{and cyclic}\,.\label{Hom''}
\end{eqnarray}
With $\mathfrak{D}^a$ and $\mathfrak{B}^a$ as contravariant vector
densities and $H_a$ and $E_a$ as ordinary covectors, this system of
equations is generally covariant, in spite of containing only partial
(and not covariant) derivatives.

Equation (\ref{Inhom''})$_2$ can be written more coherently, if we
introduce the contravariant bi-vector density
$\mathfrak{H}^{ab}:=\epsilon^{abc}H_c= -\mathfrak{H}^{ba}$;
analogously in (\ref{Inhom''})$_2$ we take
$B_{ab}=\epsilon_{abc}\mathfrak{B}^c=-B_{ba}$; here
$\epsilon_{abc}=\pm 1,0$ is the totally antisymmetric Levi-Civita
tensor density. Collecting all terms, we have (summation convention),
\begin{eqnarray}\label{Inhom'''}
  \partial_a \mathfrak{D}^a = \hat{\rho}\,,&&\qquad\hspace{11pt}
  \partial_b \mathfrak{H}^{ab} -\partial_0  \mathfrak{D}^a=\hat{j}^a\,, 
  \\
\partial_{[a} B_{bc]} =0\,,&&\qquad
\partial_{[a} E_{b]} +\frac 12\partial_0 B_{ab}=0\,,\label{Hom'''}
\end{eqnarray}
together with the Lorentz force
\begin{equation}\label{force}
F_a=q\left(E_a+B_{ab}v^b \right)\,,\qquad F_av^a=qE_a v^a\,.
\end{equation}
We have then the electric field strength $E_a$ as covector and the
magnetic field strength $B_{ab}=-B_{ba}$ as bi-covector. 

The generally covariant formula (\ref{force}) can be read as defining
operationally the electric and magnetic field strengths. Hence in
future we treat $E_a,B_{ab}$ as belonging to the primary variables of
electrodynamics. The second set, namely the electric and the magnetic
excitation $\mathfrak{D}^a,\mathfrak{H}^{ab}$, is important in the
context of formulating charge conservation, since
(\ref{Inhom'''})$_2$, upon differentiation, and substituting the time
derivative of (\ref{Inhom'''})$_1$, yields the charge conservation law
in its differential version:
\begin{equation}\label{charge}
\partial_a \hat{j}^a+\partial_0 \hat{\rho}=0\,.
\end{equation}
In this way, $\mathfrak{D}^a,\mathfrak{H}^{ab}$ can be understood as
potentials of charge and current, respectively, see \cite{TT}. We
count them also as primary field variables in electrodynamics. Lorentz
force (\ref{force}) and charge conservation (\ref{charge}) are two
{\it interfaces} between the theoretical formalism of Maxwell's
equations and experiment. Activating these interfaces makes out of a
theoretical construct a physical theory (provided the constitutive
relations are additionally specified). It is for this reason that the
field variables $E_a,B_{ab}$ and $\mathfrak{D}^a,\mathfrak{H}^{ab}$,
together with $\hat{\rho},\hat{j}^q$, in the mathematical form
specified, are measurable quantities.

This is as far as we can go unless we introduce Lorentz and Poincar\'e
transformations. The advantage of the generally covariant system
(\ref{Inhom'''}), (\ref{Hom'''}) as compared to (\ref{Inhom'}),
(\ref{Hom'}) is that in (\ref{Inhom'''}), (\ref{Hom'''}) only {\it
  partial} differentiation $\partial_a$ occurs whereas in
(\ref{Inhom'}), (\ref{Hom'}) we have the nabla operator with
components $\nabla_a=\partial_a+\Gamma_a$, wherein $\Gamma_a$ denotes
the (abbreviated) Christoffel symbols.

\section{Maxwell's equations in spacetime}

It is already clear from (\ref{Inhom'''}), (\ref{Hom'''}) that we do
not need to make a transition to Poincar\'e covariance. We can go
directly to general covariance since the 4d covariance can be read off
{}from (\ref{Inhom'''}), (\ref{Hom'''}).

\subsection{In 4d tensor calculus}

We introduce in the conventional way the 4d excitation $\check{{\cal
    H}}^{ij}=(\mathfrak{H}^{ab},\mathfrak{D}^a)$, with
$i,j,...=0,1,2,3$, the current $\check{\cal
  J}^i=(\hat{\rho},\hat{j}^a)$, and the field strength
$F_{ij}=(B_{ab},E_a)$. We find
\begin{equation}\label{4dtensorMax}
  \partial_j \check{\mathcal{H}}^{ij}=\check{\mathcal {J}}^i\,,\qquad\quad 
\partial_{[i}F_{jk]}=0\,.
\end{equation}
This scheme was known to Einstein \cite{Einstein:1916} in 1916. Among
many other texts, a lucid exposition can be found in Schr\"odinger
\cite{Schroedinger}.  Contravariant bi-vector densities can
alternatively be written as covariant bi-vectors, that is, ${\cal
  H}_{ij}=\frac 12\epsilon_{ijkl} \check{\cal H}^{ab}$, and
contravariant vector densities as ${\cal
  J}_{ijk}=\frac{1}{3!}\epsilon_{ijkl}\check{\cal J}^l={\cal
  J}_{[ijk]}$; here $\epsilon_{ijkl}=\pm 1,0$ is the totally
antisymmetric Levi-Civita tensor density. Accordingly we find the
alternative version
\begin{equation}\label{4dtensorMax'}
  \partial_{[i} {\cal H}_{jk]}= {\cal J}_{ijk}\,,\qquad\quad 
\partial_{[i}F_{jk]}=0\,.
\end{equation}
This form is particularly suited for passing over to the calculus of
exterior forms. But before doing so, we will have to look at the exact
properties of the fields ${\cal H}_{ij}$, ${\cal J}_{ijk}$, and $F_{ij}$.

In contrast to RR, we do not want that a Clifford algebra formalism
dictates us which explicit form of electrodynamics we have to take as
the valid one. We refer to experiment in order to support
operationally the appropriate form of electrodynamics. First, according
to classical mechanics, {\it force is a covector} (or covariant vector);
then, by (\ref{force}), since the charge $q$ is a scalar and the
velocity $v^a$ a (contravariant) vector, the field strength $F_{ij}$
is a conventional bi-covector in 4 dimensions.

What about ${\cal H}_{ij}$? Well, we have to be a bit careful here. In
(\ref{charge}), $\hat{j}$ and $\hat{\rho}$ are conventional {\it
  densities,} that is, they transform with $|J_3|$ (the absolute value
of the 3d Jacobian), since charge has no screw sense, see above. As a
consequence $\check{\cal J}^i$ is a 4d density with transformation
factor $|J_4|$. If we lower its indices according to ${\cal
  J}_{ijk}=\frac{1}{3!}\epsilon_{ijkl}\check{\cal J}^l$, we have to
take care of the transformation properties of $\epsilon_{ijkl}$. It is
a scalar J-density of weight $-1$, in the (adapted) language of Schouten
\cite{Schouten*}, and as such transforms with the factor
$1/J_4$. Accordingly, the factor in the transformation behavior of
${\cal J}_{ijk}$ is $\text{sign}\,J_4$. In other words, ${\cal
  J}_{ijk}$ is a {\it twisted} tri-covector and, as a consequence,
${\cal H}_{ij}$ a {\it twisted} bi-covector. Therefore, the twisted
nature of the excitation and the current in electrodynamics is a
natural consequence of the mentioned interface to charge
conservation. Of course, if we restrict the considered coordinate
transformations in an ad hoc way to those of a positive Jacobian, we
don't need to care about it. But if we opt for the most general group
under which electrodynamics is invariant, then the electric current
and the electromagnetic excitation both are twisted quantities---this
is a logical consequence of the fact that a charge carries no screw
sense.

Note that our results are consistent in the sense that the charge
integral $Q:=\int \rho$ turns out to be a scalar, exactly as the
charge features in the expression for the Lorentz force (\ref{force}).

\subsection{In 4d exterior differential form calculus}

Starting from (\ref{4dtensorMax'}), it is trivial to rewrite Maxwell's
equations in exterior form notation,\footnote{This is to be compared
  with Minkowski's symbolic representation of 1907 of Maxwell's
  equations $\text{lor} f = −s, \>\text{lor} F^\star = 0$, with the
  metric dependent differential operator lor; for details see the
  discussion in \cite{HehlMinkowski}.}
\begin{equation}\label{H-eq}
d{\mathcal H}={\mathcal J}\,,\qquad\quad dF=0\,.
\end{equation}
Here the twisted 2-form ${\cal H}=\frac 12 {\cal H}_{ij}dx^i\wedge
dx^j$ etc.; the exterior differential form calculus is presented in
Frankel \cite{Ted}, for application to electromagnetism one should
compare, for example, Lindell \cite{IsmoBook}. 

Hence eventually we found the genuine face of Maxwell's
equations. Relying on differential forms, the complete independence of
Maxwell's equations of coordinates is now manifest. And the insight
about charge conservation and the Lorentz force allowed us to
interpret ${\mathcal H}$ as {\it twisted} and $F$ as {\it untwisted}
differential 2-forms. Where RR made a mistake is apparent, they messed
up the transformation behavior of the electric charge density and
attributed to the charge a screw sense that cannot be found in nature.

Let us put this into a bit broader perspective: In formulating
electrodynamics, the basic difference between the approach of RR and
the one of us is that they take a prescribed spacetime with
orientation, metric $g_{ij}$, etc.\ and press Maxwell's equations into
that straightjacket. In contrast, we are much more careful. We may
want to put charge on a (non-orientable) M\"obius band or a Klein
bottle, for example, and we are aware that the metric represents the
gravitational potential in Einstein's theory of
gravitation. Therefore, we want to expel as many orientational and
gravitational structures as possible from the fundamental laws of
electrodynamics. That is, we subscribe to the {\em premetric} approach
of electrodynamics in the tradition of Murnaghan \cite{Murnaghan},
Kottler \cite{Kottler}, Cartan \cite{Cartan:1923}, and van Dantzig
\cite{vanDantzig}, see also
\cite{TT,Post62,Tonti:1996,Bossavit,Birkbook,Dave1,Dave2,Tonti:2009}. The
premetric Maxwell equations (\ref{H-eq}) incorporate {\it topological}
information, that is, whether certain forms are closed or exact. In
particular, they are independent of metric and connection. We want to
learn about the {\it genuine face} of Maxwell's equations, not about
the illusive ``... Many Faces of Maxwell ... Equations ..'' that at
least one of the authors of \cite{dR&R} is searching for \cite{RO}.

In the corresponding axiomatic scheme \cite{Birkbook}---for an
elementary introduction see \cite{gentle}---we make minimal
assumptions about spacetime, just a 4-dimensional manifold that we
decompose into 1+3 by means of an arbitrary normalized 4d vector $n$.
One coordinate, longitudinal to $n$, is related to the physical
dimension of time and 3 coordinates, transversal to $n$, related to
the dimension of length. Then, postulating electric charge
conservation, the form of the Lorentz force density, and magnetic flux
conservation, we arrive at what we think is the genuine
coordinate-free representation of Maxwell's equations in a
4-dimensional (4d) version, see (\ref{H-eq}). Both conservation laws
are based on counting procedures, the Lorentz force law on force
measurements known from mechanics---no measurements of time intervals
or length are involved, that is, no metric needed: This axiomatic
system is premetric.

We are even able, assuming for the vacuum a local and linear
constitutive law between electromagnetic excitation $\cal H$ and
electromagnetic field strength $F$ to {\it derive} the light
cone---and thus the metric, including its signature, up to a
factor---in the geometric optics limit \cite{HOPotsdam}, provided
birefringence is forbidden \cite{LammerH}. We also find a relation
between the Lenz rule, the sign of the energy density, and the
signature of the metric \cite{Birkbook,ItinAdP,ItinHehl}. Whereas RR a
priori put in the light cone into their spacetime picture, we get it
out from local and linear electrodynamics---giving the light cone its
proper place in a theory of electromagnetism, and not presupposing it
as an intrinsic structure of spacetime.

\subsection{In 4d Clifford calculus}

RR motivated their negative and biased attitude towards twisted forms
by their wish to reformulate electrodynamics in the Clifford bundle
language. They write in the introduction to their paper: ``...if the
charge argument is indeed correct, it seems to imply that the Clifford
bundle cannot be used to describe electromagnetism or any other
physical theory.'' In our view, this claim is unsubstantiated, see
also the work of Demers \cite{Diane} on the relation of twisted forms
with Clifford algebra.

We will not go into the details of constructing the complete
Clifford-based formulation of electrodynamics; however, we would
like to demonstrate that Maxwell's equations in vacuum can be
straightforwardly recast into the Clifford formalism without
eliminating the twisted forms. In contrast to RR, we will use the
4-dimensional covariant language throughout.

Given the 2-form of the electromagnetic field strength, $F = {\frac
  12} F_{ij}dx^i\wedge dx^j$, the corresponding Clifford field (that
is, the section of the Clifford bundle over the spacetime manifold)
reads ${\cal F} ={\frac 12} F_{ij}\gamma^{[i}\gamma^{j]}$. Let us now
apply the Dirac operator ${\cal D} = \gamma^i\nabla_i$ to  ${\cal
  F}$. With the well-known identity of Clifford algebra,
\begin{equation}
\gamma^i\gamma^{[j}\gamma^{k]} \equiv g^{ij}\gamma^k - g^{ik}\gamma^j 
+ \eta^{ijkl}\gamma_5\gamma_l\,, \quad\text{with}\quad
\eta^{ijkl}=\epsilon^{ijkl}/\sqrt{-g}\,,
\end{equation}
we immediately find
\begin{equation}
{\cal D}{\cal F} = \gamma_j\nabla_iF^{ij} + {\frac 12}\gamma_5\gamma_l
\eta^{ijkl}\nabla_iF_{jk}. 
\end{equation}
For the electric current $J^i = (\rho, {\mathbf j})$ we define in the
usual way the Clifford field ${\cal J} = \gamma_i J^i$. Then we can
verify that the Clifford-algebra equation of the Dirac type
\begin{equation}
{\cal D}{\cal F} = -{\cal J}\label{MaxClifford}
\end{equation}
is completely equivalent to Maxwell's inhomogeneous and homogeneous
equations in vacuum:
\begin{equation}\label{Max4D}
  \partial_j\check{\mathcal{H}}^{ij}=\check{\mathcal{J}}^i\,,\qquad \partial_i
  F_{jk}+\partial_j F_{ki}+ \partial_k F_{ij}=0\,.
\end{equation}
Here $\check{\mathcal{J}}^i$ is the current density and the vacuum
constitutive relation is assumed to be $\check{\mathcal{H}}^{ij} =
\sqrt{-g}g^{ik}g^{jl}F_{kl}$.

Note that the Maxwell equations (\ref{Max4D}) are in their standard
form, and the electromagnetic excitation $\check{\mathcal{H}}^{ij}$ as
well as the electric current density $\check{\mathcal{J}}^i$ are both
{\it twisted}. This fact presents absolutely no difficulty to the {\it
  equivalent} Clifford-algebra formulation specified in equation
(\ref{MaxClifford}).

As a final remark we feel it necessary to stress that although we
admit that there is a certain beauty in the Clifford-algebra approach,
the latter seems to be strongly confined to {\it vacuum}
electrodynamics, and the equation (\ref{MaxClifford}) cannot be
satisfactorily extended to more general constitutive laws, except for
the case of a moving isotropic medium, see Jancewicz \cite{Bernard}.

\section{Discussion}

In our quasi-historical description we wanted to show in a simple
manner how the concepts developed over time for the description of
electrodynamics: from 3-dimensional Cartesian vector calculus to the
differential form presentation in (\ref{H-eq}). There are alternative,
and perhaps still more convincing approaches by starting from discrete
electrodynamics and using the theory of chains and cochains, see
Bossavit \cite{Bossavit}, Tonti \cite{Tonti:1996,Tonti:2009},
Zirnbauer \cite{Zirnbauer}, and others. In the end, in a continuum
limit, these authors found the same structures as in (\ref{H-eq}), in
particular, they also found unequivocally twisted forms. Classical
electrodynamics is a closely knit structure and one cannot introduce
changes at one structure without affecting badly other structures.
                                    
As we argued, RR did not recognize the importance of the interfaces
between mathematical theory and experiment. This can also be seen in
another way. In the Abstract RR claimed that ``... we derive directly
from Maxwell equation the density of [the Lorentz] force ...'' How
come? RR started from Maxwell's equations (see their Sec.~6) in which
there occurs a mathematical symbol $F$ without physical meaning; at
least they did not specify an operational definition of $F$, that is,
how one has to measure it. Then RR manipulated the Maxwell's equations
in the conventional way and came up with the energy-momentum tensor of
the electromagnetic field in vacuum, which they assumed to be
known---besides Maxwell's equations. For the divergence of the
energy-momentum tensor they found [if we translate it into the
notation of our equations in (\ref{H-eq})] the force density
$f_\a=(e_\a\ F)\wedge {\cal J}$. In this formula, the force density
$f_\a$ is linked to the unidentified freely floating object $F$ and
the electric current $\cal J$. {\em Provided $F$ is identified with the
electromagnetic field strength}---and this is what RR did---this
formula is the expression for the Lorentz force density. Consequently,
RR identified their $F$ as field strength by means of the Lorentz formula.

Now, they claimed that two of us \cite{Birkbook} took the Lorentz
force density as an axiom, but they {\it derived} it. This is an empty
claim because their $F$ was an unidentified object and nothing
else. RR found out eventually that $F$ can be be identified as the
electromagnetic field strength via the Lorentz formula, whereas we
took it as an axiom. RR did not recognize that {\it logically} they
did the same as we did; but we, for reasons of transparency and
straightforwardness, formulated our assumptions at the beginning
within our axiomatic scheme whereas RR made the same assumption in a
hidden way at the end of their calculations.

The experience one had won from the classical electrostatic
experiments of the 18th and 19th centuries was that electric charge
$Q:=\int_{\Omega_3}\rho$ is a 3d scalar. In modern elementary particle
physics, the conservation of electric charge is a well-tested
law. Consider a scattering process of an electron with a neutron (or a
proton). One ascribes to the charge of each particle a number and adds
up these numbers before scattering and afterwards: The electron
carries a negative elementary charge $-1$, the neutron or the proton
carry a $0$ or a $+1$, respectively. There is no screw, chirality, or
handedness involved in testing charge conservation. {\em Scalar
  numbers} and adding up them is all what is required.

This confirms the conclusion that the charge also in classical
electrodynamics is a pure scalar. Any other attribute to the charge
would not be reflected in nature, would be superfluous and
redundant. This conclusion is consistent with Schouten's verdict:
``... an electric charge has no screw sense.'' See \cite{Schouten*},
p.\ 132. 

But doesn't carry an electron, the carrier of an elementary charge, a
spin? Isn't then a screw attached to it? Yes, indeed, but this screw
is exclusively related to spin angular momentum of the electron, but
not to its charge. Take a negatively charged pion $\pi^-$. It carries
no spin, but an electric charge---and in this case there is no screw
related to the $\pi^-$.

Accordingly, experiments show convincingly that the electric charge
has no screw attached to it.  Therefore, in formalizing charge
conservation, the 4-dimensional electric current $\cal J$ has to be a
3-form {\em with} twist. Only then the charge $Q=\int_{\Omega_3} {\cal
  J}$ is really a 4-dimensional scalar, totally independent of any
orientation of spacetime. As Perlick \cite{Perlick} remarked so aptly:
``... one {\it must} understand the excitation as a form with twist if
one wants that the charge contained in a volume always has the same
sign, independent of the orientation chosen.'' See also Bossavit
\cite{Bossavit} and Tonti \cite{Tonti:1996,Tonti:2009} in this
context.  This is in marked contrast to the RR-formalism: therein the
charge $Q$ switches its sign upon turning around the orientation. They
try to fix it by additional {\it ad hoc} assumptions, but it is
evident: Their {\em surrogate charge} carries an additional attribute
that has no image in nature. Their `charge' is over-freighted with a
redundant structure.
\bigskip

\noindent{\bf Note added in proof:} In a `note added in proof' in
\cite{dR&R}, da~Rocha and Rodrigues tried to answer our above
formulated objections to their article \cite{dR&R}, see also their
paper \cite{RRRR}. We will discuss
shortly their main points in the sequence chosen by RR:

\begin{enumerate}

\item Twistfree electrodynamics and ``observed phenomena''

  We agree with RR that one can formulate a twistfree electrodynamics,
  that is, using only untwisted forms. All what we tried to point out
  is that this amounts to an amputation: essential properties of
  electrodynamics are cut-off (see the example of the M\"obius strip
  below). The quoted lemma of de Rhams is a purely geometrical
  statement, notions such as integrals over forms, like an action, are
  not discussed, nor observed parity violations like the one in the
  electrodynamics of the antiferromagnet Cr$_2$O$_3$
  \cite{SantaBarbara}. That charge carries no screw is---in contrast
  to RR's statement to the opposite---an experimentally esablished
  fact, we mentioned the {\it scattering experiments} in Sec.3 above.

\item Charge on a M\"obius strip and twisted differential forms

  Recall that a twisted form is a special mathematical construct that
  gives a {\it well-defined} value to an integral over a certain
  domain.  This value is independent of the orientation and even of
  the orientability of the domain.  In particular, twisted forms
  provide positive values for length, volume, and mass-energy.

  Also the total electric charge turns out to have a {\it
    well-defined} (positive or negative) value. RR \cite{dR&R}
  claimed: ``Now had our critics read our Remark 13 they could be
  recalled of the fact that being $\cal J$ a pair {\em or} an impair
  2-form we cannot define its integral over the M\"obius strip. So, we
  conclude that only in fiction can someone think in putting a real
  physical charge distribution (made of elementary charge carriers) on
  a M\"obius strip ... and leaving this physical impossiblity aside we
  cannot see any necessity for the use of impair [twisted] forms.''
  We disagree with this inverse logic strongly. One cannot {\it
    conclude} anything about real physics from their specific
  mathematical constructions. The only possibility for them is {\it to
    claim} some physics behavior based on their ``good mathematics''
  and to compare it to real physics.  

  The facts are: An electrically conducting M\"obius strip can be
  constructed and electrically charged, see Stewart
  \cite{Stewart}. Its one sidedness can be observed by electrostatic
  means even in a simple experiment in a school laboratory
  \cite{Stewart}. Accordingly, who is talking about `fiction' and a
  `physical impossibility'? Perhaps it is safer to adhere to twisted
  forms in order to be able, unlike RR, to integrate the charge over a
  M\"obius strip.

\item Clifford bundle is consistent with twisted forms

  By their remark concerning the Clifford bundle approach, RR
  introduced nothing but confusion. Contrary to the original claim
  (see their introduction) that the twisted forms and axial vectors
  are incompatible with the Clifford approach, now they seem to agree
  with the opposite, namely, that it is still possible to keep working
  with the twisted forms along with the Clifford structures. Fine!
  This is consistent with what we said. As soon as it is perfectly
  possible to live happily with the axial vectors and twisted forms
  within the Clifford bundle framework, one should be strongly advised
  to keep using them. It is thus satisfactory to see that RR agree
  with our conclusion that their approach, in which the well-defined
  charge, mass, and volume are replaced with the orientation-dependent
  surrogates, is unwarranted and redundant.

\item Lorentz force used for an operational definition of the field
  strength

  We quote RR: ``... we proved that the coupling of $F$ with $J$ must
  be given by the Lorentz force law, which must then be used in the
  {\it operational} way in which those objects must be used when one is
  doing Physics.''

  In other words, the Lorentz force formula must be used operationally
  in order to define the meaning of $F$. This is exactly what we
  claimed. And this operational interpretation is a coditio sine qua
  non. Whether one does it at the beginning or at the end of an
  electrodynamic theory doesn't make a logical difference.

\end{enumerate}

\begin{acknowledgement}
  We are very grateful to Waldyr Rodrigues for many interesting email
  exchanges on the subject of our joint interest. Moreover, we thank
  Volker Perlick (Lancaster) for useful discussions.
\end{acknowledgement}


\begin{thebibliography}{[00]}

\bibitem{dR&R} R.~da~Rocha and W.~A.~Rodrigues jr., {\em Pair and
    impair, even and odd form fields, and electromagnetism,}
  Ann.\ Phys.\ (Berlin), to be published (2010) [arXiv:0811.1713v7
  [math-ph]].

\bibitem{Maxwell} J.~C.~Maxwell, {\it A dynamical theory of the
    electromagnetic field,} see \cite{Niven} Vol.1, pp.\ 526--596, for
  Maxwell's equation, see particularly pp.\ 554--564 (1864).

\bibitem{Niven} W.~D.~Niven, editor, {\it The Scientific Papers of
    James Clerk Maxwell,} Two volumes bound as one (Dover, New York,
  1965).

\bibitem{Abraham} M.~Abraham and A.~F\"oppl, {\it Theorie der
    Elektrizit\"at I,} 2nd ed. (Teubner, Lepzig, 1904).

\bibitem{Jackson} J.~D.~Jackson, {\it Classical Electrodynamics,} 3rd
  ed. (Wiley, New York, 1999).

\bibitem{Gelman} H.~Gelman, {\it Handed Products and Axial Vectors,}
  Am.\ J.\ Phys.\ {\bf 38}, 599--608 (1970).

\bibitem{Brevik} I.~Brevik, {\it Polar and Axial Vectors in
    Electrodynamics,} Am.\ J.\ Phys.\ {\bf 40}, 550--552 (1972).

\bibitem{Yavari} A.~Yavari, {\it On geometric discretization of
    elasticity}, {J.\ Math.\ Phys.} {\bf 49}, 022901 [36 pages] (2008).

\bibitem{Schouten*} J.~A.~Schouten, {\it Tensor Analysis for
    Physicists,} 2nd ed.\ reprinted (Dover, Mineola, New York, 1989).

\bibitem{TT} C.~Truesdell and R.~A.~Toupin, {\it The classical field
    theories}, in: {\sl Handbuch der Physik}, Vol.\ III/1, edited by
  S.~Fl{\"u}gge (Springer, Berlin, 1960) pp.\ 226--793.

\bibitem{Einstein:1916} A.~Einstein, {\it Eine neue formale Deutung
    der Maxwellschen Feld\-glei\-chungen der Elektrodynamik}, {
    Sitzungsber.\ K\"onigl.\ Preuss.\ Akad.\ Wiss.\ Berlin} (1916)
  pp.\ 184--188; see also {\em The collected papers of Albert Einstein.}
  Vol.~6, A.~J.~Kox et al., eds. (Princeton University Press, Princeton
  NJ, 1996) pp. 263--269.
 
\bibitem{Schroedinger} E.~Schr\"odinger, {\it Space-Time Structure}
  (Cambridge University Press, Cambridge, 1954).

\bibitem{HehlMinkowski} F.~W.~Hehl, {\it Maxwell's equations in
    Minkowski's world: their premetric generalization and the
    electromagnetic energy-momentum tensor,} Ann.\ Phys.\ (Berlin)
  {\bf 17}, 691--704 (2008) [arXiv:0807.4249].
 
\bibitem{Ted} T.\ Frankel, {\it The Geometry of Physics} -- An
  Introduction (Cambridge University Press, Cambridge, 1997, 1999).

\bibitem{IsmoBook} I.~V.~Lindell, {\it Differential Forms in
    Electromagnetics} (IEEE Press, Piscataway, NJ, and
  Wiley-Interscience, 2004).

\bibitem{Murnaghan} F.~D.~Murnaghan, {\it The absolute significance of
    Maxwell's equations,} Phys.\ Rev.\ {\bf 17}, 73--88 (1921).

\bibitem{Kottler} F.~Kottler, {\it Maxwell'sche Gleichungen und
    Metrik}, {Sitz\-ungs\-ber.\ Akad.\ Wien IIa} {\bf 131}, 119--146
  (1922).

\bibitem{Cartan:1923} \'E.\ Cartan, {\it On
    Manifolds with an Affine Connection and the Theory of General
    Relativity}, English translation of the French original of 1923/24
  (Bibliopolis, Napoli, 1986).

\bibitem{vanDantzig} D.\ van Dantzig, {\it The fundamental equations
    of electromagnetism, independent of metrical geometry}, {\sl Proc.
    Cambridge Phil. Soc.} {\bf 30}, 421--427 (1934).

\bibitem{Post62} E.~J.~Post, {\it Formal Structure of Electromagnetics
    -- General Covariance and Electromagnetics} (North Holland,
  Amsterdam, 1962, and Dover, Mineola, New York, 1997).

\bibitem{Tonti:1996} E.~Tonti, {\it On the geometrical structure of
    electromagnetism,} in: {\it Gravitation, Electromagnetism and
    Geometrical Structures,} Festschrift for Lichnerowicz, edited by
  G.\ Ferrarese (Pitagora Editrice, Bologna, 1996) pp.\ 281--308.

\bibitem{Bossavit} A.~Bossavit, {\it On the geometry of
    Electromagnetism,} particularly {\it The Maxwell House}, see on
  his homepage http://butler.cc.tut.fi/$\tilde{\hspace{8pt}}$bossavit/
  under ``Japanese Papers'' (1998--2000).

\bibitem{Birkbook} F.~W.~Hehl and Yu.~N.~Obukhov, {\it Foundations of
    Classical Electrodynamics: Charge, Flux, and Metric}
  (Birkh\"auser, Boston, MA, 2003).
      
\bibitem{Dave1} D.~H.~Delphenich, {\it On the axioms of topological
    electromagnetism,} Ann.\ Phys.\ (Berlin) {\bf 14}, 347--377
  (2005) [arXiv.org/hep-th/0311256].

\bibitem{Dave2} D.~H.~Delphenich, {\it Pre-metric Electromagnetism,}
  unpublished manuscript [410 pages] (May 2009).

\bibitem{Tonti:2009} E.~Tonti, {\it On the geometrical structure of
    electromagnetism,} Preprint November 2009 [50 pages], see also
  Tonti's homepage with references to earlier papers:
  http://www.dic.univ.trieste.it/perspage/tonti/ .

\bibitem{RO} W.~A.~Rodrigues jr. and E.~C.~de~Olivera, {\it The Many
    Faces of Maxwell, Dirac and Einstein Equations, a Clifford bundle
    approach}, Lect.\ Notes in Phys.\ {\bf 722} (Springer, Berlin,
  2007) pp.\ 1--445.

\bibitem{gentle} F.~W.\ Hehl and Yu.~N.\ Obukhov, {\em A gentle
    introduction into the foundations of classical electro\-dynamics:
    Meaning of the excitations $({\mathcal D},{\mathcal H})$ and the
    field strengths $(E,B)$} [arXiv: physics/0005084] (2000).
    
\bibitem{HOPotsdam} F.~W. Hehl and Yu.~N. Obukhov, {\it Spacetime
    metric {}from local and linear electrodynamics: A new axiomatic
    scheme}, in: {\sl ``Special relativity: Will it survive the next
    101 years?''}, Proc. of the 271st Heraeus-Seminar, 13-18 February
  2005, Potsdam.  Edited by J.~Ehlers and C.~L\"ammerzahl, {Lect.\
    Notes in Phys.}  {\bf 702} (Springer, Berlin, 2006) pp.\ 163--187
  [arXiv.org/gr-qc/0508024].

\bibitem{LammerH} C.~L\"ammerzahl and F.~W.~Hehl, {\it Riemannian
    light cone from vanishing birefringence in premetric vacuum
    electrodynamics,} Phys.\ Rev.\ D {\bf 70}, 105022 (2004) (7 pages)
  [arXiv.org/gr-qc/0409072].
            
\bibitem{ItinAdP} Y.~Itin and Y.~Friedman, {\it Backwards on
    Minkowski's road. From $4D$ to $3D$ Maxwellian electromagnetism,}
  Ann.\ Phys.\ (Leipzig) {\bf 17}, 769-–786 (2008).

\bibitem{ItinHehl} Y.~Itin and F.~W.~Hehl, {\it Is the Lorentz
    signature of the metric of spacetime electromagnetic in origin?}
  Ann.\ Phys.\ (NY) {\bf 312}, 60--83 (2004)
  [arXiv.org/gr-qc/0401016].

\bibitem{Diane} D.~G.~Demers, {\it Nonassociative, Clifford-like
    algebra}, http://felicity.freeshell.org/math/ and links specified
  there; see also her abstract under
  http://felicity.freeshell.org/math/ExtCalc.pdf .

\bibitem{Bernard} B.~Jancewicz, {\it Multivectors and Clifford Algebra
    in Electrodynamics} (World Scientific, Singapore, 1988).
%
%
\bibitem{Zirnbauer} M.~Zirnbauer, {\it Maxwell in chains,} colloquia
  given at different places, unpublished manuscript (Cologne 2001);
  see also his homepage http://www.thp.uni-koeln.de/zirn/index.html .

\bibitem{Perlick} V.~Perlick, private communication (March 2007).

\bibitem{SantaBarbara} F.~W.~Hehl, Y.~N.~Obukhov, J.-P.~Rivera and
  H.~Schmid, {\it Magnetoelectric Cr$_2$O$_3$ and relativity theory,}
  Eur.\ Phys.\ J.\ B {\bf 71}, 321-–329 (2009) [arXiv:0903.1261].

\bibitem{Stewart} S.~Stewart, {\it Flexible Faraday cage with a twist:
    surface charge on a M\"obius strip,} Phys.\ Teacher {\bf 45},
  268--269 (2007).

\bibitem{RRRR} R.~da Rocha and W.~A.~Rodrigues, arXiv:0912.2127v1
  [math-ph].

\end{thebibliography}
\end{document}